\documentclass[12pt]{article}
\pdfoutput=1

\usepackage{draft,hyperref,tikz,float,subfig}

\usetikzlibrary{snakes}
\usetikzlibrary{shapes.misc}

\newcommand{\mf}{\mathfrak}
\newcommand{\mZ}{\mathbb Z}
\newcommand{\mR}{\mathbb R}

\newcommand\nn{\nonumber}
\def\({\left(}
\def\){\right)}

\begin{document}

	\begin{titlepage}
		
			 \begin{flushright}
		\hfill{\tt CALT-TH 2020-002, IPMU20-0023, PUPT-2609}
		\end{flushright}
		
		\begin{center}
			
			\hfill \\
			\hfill \\
			\vskip 1cm
			
			\title{Twist Gap and Global Symmetry \\ in Two Dimensions
			}
			
			\author{Nathan Benjamin,$^{a}$ Hirosi Ooguri,$^{b,c}$  Shu-Heng Shao,$^d$ and Yifan Wang$^{e,f}$
			}
			
			\address{
				${}^a$Princeton Center for Theoretical Science, \\ Princeton University, 
				Princeton, NJ 08544, USA
				\\
				${}^b$Walter Burke Institute for Theoretical Physics,  \\
				California Institute of Technology, Pasadena, CA 91125, USA
				\\
				${}^c$ Kavli Institute for the Physics and Mathematics of the Universe (WPI),  \\
				University of Tokyo, Kashiwa, 277-8583, Japan
				\\
				${}^d$School of Natural Sciences, Institute for Advanced Study, 
				Princeton, NJ 08540, USA
				\\
				${}^e$Center of Mathematical Sciences and Applications, \\ Harvard University,  
				Cambridge, MA 02138, USA
				\\
				${}^f$Jefferson Physical Laboratory, Harvard University, 
				Cambridge, MA 02138, USA
			}
			
			\email{nathanb@princeton.edu, ooguri@caltech.edu, \\ shao@ias.edu, yifanw@g.harvard.edu}
			
		\end{center}

		\vfill
		
		\abstract{
			We show that every compact, unitary two-dimensional CFT with    an abelian conserved current    has vanishing twist gap for charged primary fields with respect to the $\mf{u}(1)\times$Virasoro algebra.  
  This means that either the chiral algebra is enhanced by a charged primary field with zero twist 
or there is an infinite family of charged primary fields  that accumulate to zero twist. 
		}
		
		\vfill

	\end{titlepage}
	
	\eject
	
	\tableofcontents
	
	\unitlength = .8mm

	\section{Introduction}
	
The study of two-dimensional conformal field theories (CFTs) has a long history in theoretical physics. Although much work has been done in understanding rational conformal field theories (RCFTs),\footnote{Here we take the definition of RCFTs as 2d CFTs with a finite number of conformal blocks under a certain chiral algebra.} there is very little general understanding of a generic, irrational 2d CFT. One feature  that can diagnose irrationality of a CFT is the \emph{twist gap} of the theory with respect to a certain chiral algebra. The twist, $2t_{\mathcal{O}}$, of an operator $\mathcal{O}$ with total conformal dimension $\Delta$ and spin $j$ is defined as
\be
2t_{\mathcal{O}} = \Delta - |j| = 2\text{min}(h, \bar h),
\ee
where $h$ and $\bar h$ are the left- and right-moving conformal dimensions of the operator.
The twist gap of a theory is the minimum (or infimum) twist of all non-vacuum primary operators under the chiral algebra. For instance, if a CFT has a conserved current as a primary operator with respect to a certain chiral algebra, the twist gap of that CFT vanishes under that chiral algebra. Given any chiral algebra, all rational CFTs must have vanishing twist gap under that chiral algebra for sufficiently large central charge. To be precise, if the vacuum Verma module of a chiral algebra grows   asymptotically as a $c=c_*$ CFT, then all RCFTs with $c>c_*$ must have extra currents (and therefore vanishing twist gap) under that chiral algebra\footnote{To be more precise, we define $c_*$ of a chiral algebra as the following. We say the number of vacuum descendants at dimension $\Delta$ of the chiral algebra goes as $e^{2\pi\sqrt{\frac{c_* \Delta}6}}$ at large $\Delta$, up to sub-exponential corrections. For $c>c_*$, the Cardy formula \cite{Cardy:1986ie} shows the number of primary operators under the chiral algebra goes as $e^{2\pi\sqrt{\frac{(c-c_*)\Delta}3}}$. Thus if $c>c_*$ there   must be infinitely many primary operators under this chiral algebra, which means that the RCFT must have additional currents.}. Moreover, if a CFT has exactly marginal operators, and RCFTs are dense in the conformal manifold (for instance, in the case of the moduli space of the $c=1$ free boson which we discuss more below), then the twist gap under the chiral algebra shared by all theories in the moduli space must vanish everywhere.

In Section 2.2 of \cite{Collier:2016cls}, a rigorous upper bound was placed on the twist gap of any $c>1$ compact, unitary CFT under the Virasoro algebra\footnote{In \cite{Collier:2016cls}, the argument was credited to Tom Hartman.}. It was shown that 
\be
2t_{\text{gap}}^{\text{Virasoro}} \leq \frac{c-1}{12}.
\ee
In \cite{Benjamin:2019stq}, it was argued that this bound can be improved to
\be
2t_{\text{gap}}^{\text{Virasoro}} \leq \frac{c-1}{16} 
\ee
from positivity of the spectrum of the theory. We emphasize that both of these bounds are nonzero, and indeed it is expected that a generic interacting CFT has a finite, nonzero twist gap under the Virasoro algebra (although we pause to note that we do not know of any explicit example of such a theory). The latter improved twist gap bound was recently revisited in \cite{Alday:2019vdr} using a Rademacher expansion of the torus partition function. 

In this paper we will show that in any compact CFT with an abelian conserved current $(J,\bar J)$ whose holomorphic Fourier modes $J_n$ generate the $\mf{u}(1)$ chiral algebra  (similarly for the anti-holomorphic part),\footnote{We use  $\mf{u}(1)$ for a rank-one
abelian Lie algebra, and we reserve $U(1)$ for a compact abelian Lie group. 
When the holomorphic current $J$ has a preferred normalization (e.g. when its charges are quantized), we will denote the corresponding chiral algebra by $\mf{u}(1)_k$ where $k$ is the  level of the current algebra,  and similarly for the anti-holomorphic current $\bar J$.} the twist gap under the $\mf{u}(1)\times$Virasoro algebra vanishes:
\be
2t_{\text{gap}}^{\mf{u}(1)\times\text{Virasoro}} = 0.
\ee
Thus the situation for the $\mf{u}(1)$ current algebra is in stark contrast to the Virasoro algebra. 

Note that the holomorphic current $J$ (or the anti-holomorphic component $\bar J$) generates either a compact $U(1)$ or $\mR$ symmetry, and we do not make assumptions on this in the paper.
In other words, we show that there is always a charged primary operator under the $\mf{u}(1)$ current algebra that is either a conserved (higher-spin) current, or an infinite family of charged primary operators that accumulate to zero twist. 

The prototypical example is the $c=1$ compact boson at radius $R$.  
When $R^2$ is rational, there is a holomorphic current extending the $\mf{u}(1)\times$Virasoro algebra, and the twist gap with respect to the $\mf{u}(1)\times$Virasoro algebra  vanishes.  
On the other hand, when $R^2$ is irrational, there is a tower of (non-holomorphic) primary operators that asymptote to vanishing twist. 

As an application, we also show that any $c>3$ CFT with $\mathcal{N}=2$ superconformal symmetry has a vanishing twist gap under the (unextended) $\mathcal{N}=2$ super-Virasoro algebra. 
	
	\section{$U(1)$ Global Symmetry}
	\label{sec:global}
	
	In this section, we discuss properties of 2d CFTs with compact $U(1)$ global symmetry
(the compactness assumption will be relaxed in the next section).

	\subsection{Currents and Charges}
Consider a 2d compact, unitary, bosonic CFT with 
a $U(1)$ global symmetry.  Being a  bosonic CFT, all local operators have
 integral Lorentz spin, $h-\bar h\in \bZ$.  
	We will later generalize our arguments for fermionic CFTs. 
	It will be important that the global structure of the symmetry is $U(1)$ not $\mathbb{R}$.

	Consider a $U(1)$ global symmetry generated by the current $J_\mu(z,\bar z)$ that acts faithfully on the Hilbert space.   Let the holomorphic and antiholomorphic components of the current be $J(z)\equiv J_z$ and $\bar J(\bar z)\equiv J_{\bar z}$, respectively.  
	In any compact unitary two-dimensional  CFT,  current conservation and unitarity imply $\partial \bar J= 0$ and $\bar\partial J=0$.  However, the holomorphic current $J(z)$ alone may not necessarily generate a compact $U(1)$ group, 
but $\mathbb{R}$ rather, and similarly for the anti-holomorphic current $\bar J$.

	The conserved charge  $U_\eta$ with $\eta\in [0,1)$ for this $U(1)$ global symmetry is supported on a curve $L$ in the Euclidean spacetime:
	\begin{align}
		U_\eta = :\exp \left[ \eta\left(  \oint_L dz J(z) - \oint _L d\bar z \bar J(\bar z) \right)\right] :\,,
	\end{align}
	where $ :$ stands for normal-ordering. 	Current conservation $\partial \bar J=\bar\partial J=0$ implies that $U_\eta$ is invariant under small deformation of the curve $L$.  
	Hence $U_\eta$ is a topological line operator.  
	See, for example, \cite{Bhardwaj:2017xup,Chang:2018iay,Lin:2019kpn} for discussions on topological lines in two dimensions.   
	The $U(1)$ condition implies that $\eta$ is circle valued, i.e.\ $\eta\sim \eta+1$.  In particular, $U_{\eta=1} = I$ is the identity operator on the Hilbert space. The faithfulness condition implies that $U_\eta$ is not an identity operator unless $\eta$ is an integer.
	
	The OPEs of $J$ and $\bar J$ are
	\begin{align}
		&J(z) J(0) \sim {k \over z^2}\,,~~~~~\bar J(\bar z) \bar J(0) \sim {\bar k\over \bar z^2}\,.
	\end{align}
	Having specified a faithful $U(1)$ global symmetry, the levels $k$ and $\bar k$ are physically meaningful and cannot be scaled away.\footnote{For example, if we had rescaled both $J\to 2J$ and $\bar J \to 2\bar J$, then the new topological line operator $U_\eta$ with $\eta=1/2$ would act trivially on the Hilbert space $\cal H$, violating the faithfulness condition.}

	\subsection{Spectral Flow and the Twisted Hilbert Space}\label{sec:spectralflow}
	
\begin{figure}
	\centering
	\includegraphics[width=.6\textwidth]{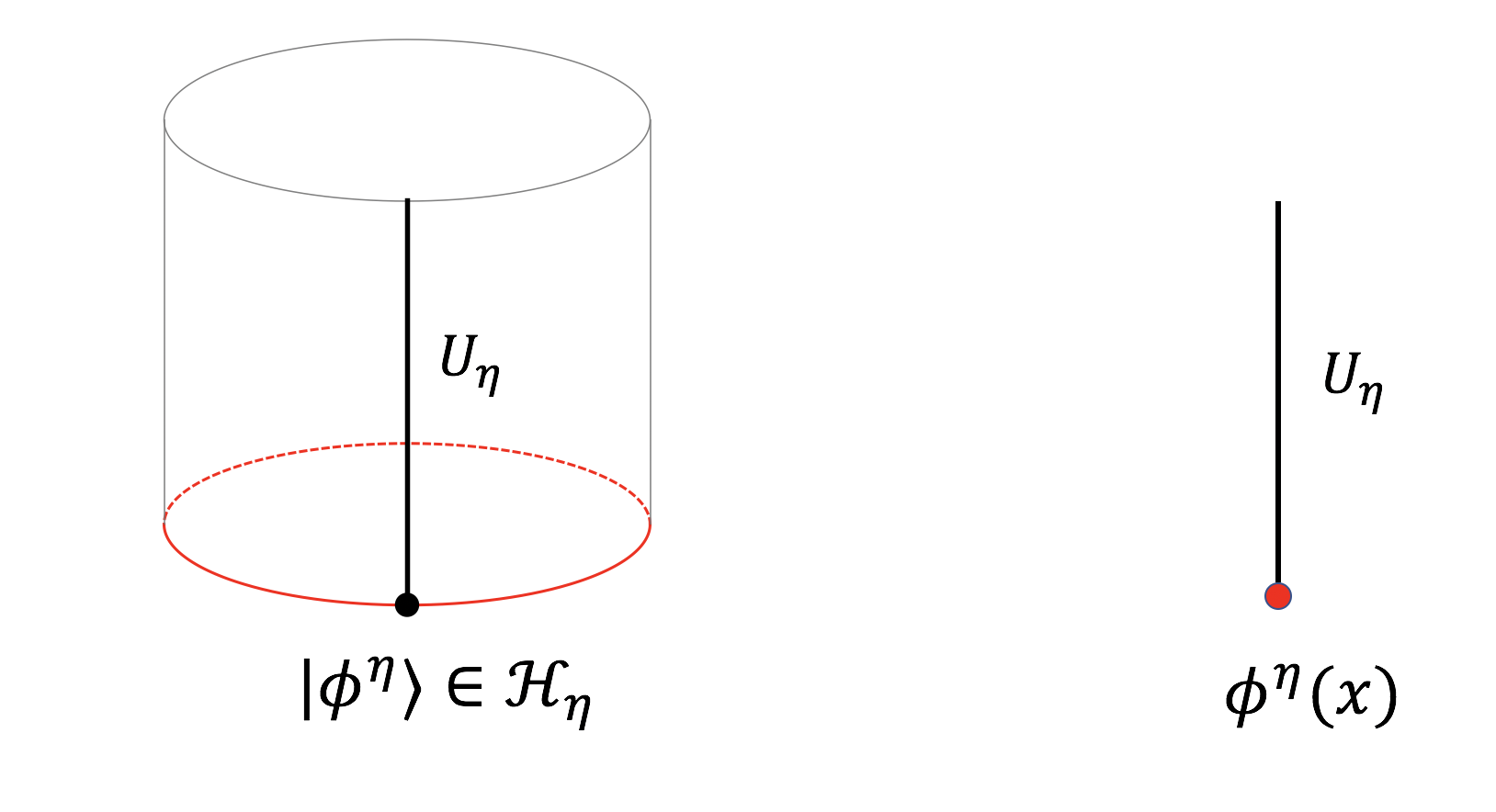}
	\caption{Using the operator-state correspondence, a state $|\phi^\eta\rangle$ in the twisted Hilbert space ${\cal H}_\eta$ is mapped to a non-local, point-like operator attached to a topological line defect $U_\eta$.}\label{fig:opstate}
	\end{figure}

	When $U_\eta$ is supported on the whole space at a fixed time, it is an operator acting on the Hilbert space $\cal H$.  
	On the other hand, when $U_\eta$ is supported at a fixed position in space but extends in time, it is  a defect.  
	The insertion of this defect gives rise to a twisted boundary condition for the quantization.  
	We will denote the Hilbert space on a circle $S^1$ with an insertion of a defect $U_\eta$ as ${\cal H}_\eta$. 
	Via the operator-state correspondence, a state in ${\cal H}_\eta$ is mapped to a \textit{non-local}, point-like operator living at the end of the topological line $U_\eta$ on the plane (see Figure \ref{fig:opstate}). 
	In particular, ${\cal H} \equiv {\cal H}_{\eta=0}$ is the original Hilbert space of local operators. 
	
	What is the relation between the twisted Hilbert space ${\cal H}_\eta$ and the Hilbert space of local operators $\cal H$? 
	This is explained by the spectral flow, which is an isomorphism between ${\cal H}_\eta$ and $\cal H$ \cite{Lin:2019kpn}. 
	More precisely,  spectral flow maps a state $|\phi\rangle \in\cal H$ with quantum numbers $(L_0,\bar L_0,J_0,\bar J_0)$ to a state in $|\phi^\eta\rangle\in {\cal H}_\eta$ with quantum numbers $(L_0^\eta,\bar L_0^\eta,J_0^\eta,\bar J_0^\eta)$ \cite{Schwimmer:1986mf}: 
	\begin{align}\label{sf}
		\begin{split}
			&L_0^\eta = L_0 -\eta J_0  + {k\eta^2\over2}\,,~~~J_0^\eta = J_0 -\eta k \,,\\
			&\bar L_0^\eta = \bar L_0 + \eta \bar J_0  + {\bar k\eta^2\over2}\,,~~~\bar J_0^\eta =\bar J_0 +\eta \bar k \,.
		\end{split}
	\end{align}

	Spectral flow is related to the modular property of the partition function, as we demonstrate in the following. 
	Consider the torus partition function  with general chemical potentials for the  symmetry generators $J_0$ and $\bar J_0$:
		\begin{align}\label{H}
		Z(\tau,  \bar \tau , \eta , \bar \eta)
		&= \text{Tr}_{\cal H} \, \exp\left[{2\pi i \tau } ( L_0 -{c\over 24} ) 
		-{2\pi i \bar \tau} (\bar L_0 -{c\over 24})
		+2\pi i \eta J_0 -  2\pi i \bar\eta \bar J_0
		\right] \,.
	\end{align}
	The torus partition function satisfies the modular property \cite{Benjamin:2016fhe}
	\begin{align}\label{holoS}
		Z(-1/\tau  , -1/\bar \tau , \eta /\tau  , \bar\eta/\bar \tau) = \exp \left( ik  \pi {\, \eta^2\over \tau }
		- i\bar k  \pi \, {\bar\eta^2\over \bar\tau}
		\right)
		Z(\tau,\bar\tau ,\eta,\bar \eta) \,.
	\end{align}
	Let us understand this modular property in the special case when $\bar\eta= - \eta$. 
	In this case, the partition function $Z(\tau,\bar\tau,\eta)$ on the righthand side is the trace over the  Hilbert space $\cal H$ with $U_\eta$ inserted at a fixed time:
	\ie
	Z(\tau,\bar \tau,\eta,\bar \eta=-\eta) = \text{Tr}_{\cal H} \,\left\{ U_\eta \,  \exp\left[ 2\pi i\tau  (L_0^\eta-{c\over24} )-{2\pi i \bar\tau}  (\bar L_0^\eta-{c\over24} )  \right]\right\}\,.
	\fe
	It also admits an $S$-dual interpretation as the partition function over the twisted Hilbert space ${\cal H}_\eta$ on a torus with modulus $-1/\tau$:
	\ie\label{zr}
	Z(\tau,\bar \tau,\eta,\bar \eta=-\eta) = \text{Tr}_{{\cal H}_\eta} \exp\left[ -{2\pi i\over \tau}  (L_0^\eta-{c\over24} )+{2\pi i\over \bar\tau}  (\bar L_0^\eta-{c\over24} )  \right]\,.
	\fe
	The lefthand side of \eqref{holoS}, on the other hand, can be written as a sum over the original Hilbert space:
	\ie\label{zl}
	Z(-1/\tau  , -1/\bar \tau , \eta /\tau  ,-\eta/\bar\tau)  = \text{Tr}_{\cal H} \exp\left[ -{2\pi i\over \tau}  (L_0-{c\over24} )+{2\pi i\over \bar\tau}  (\bar L_0-{c\over24} )    + {\eta\over \tau} J_0 +  {\eta\over\bar\tau} \bar J_0)\right]\,.
	\fe
	Using the expressions \eqref{zr} and \eqref{zl}, we see that \eqref{holoS} is reproduced by the spectral flow map \eqref{sf}.

\begin{figure}
	\centering
	\includegraphics[width=.75\textwidth]{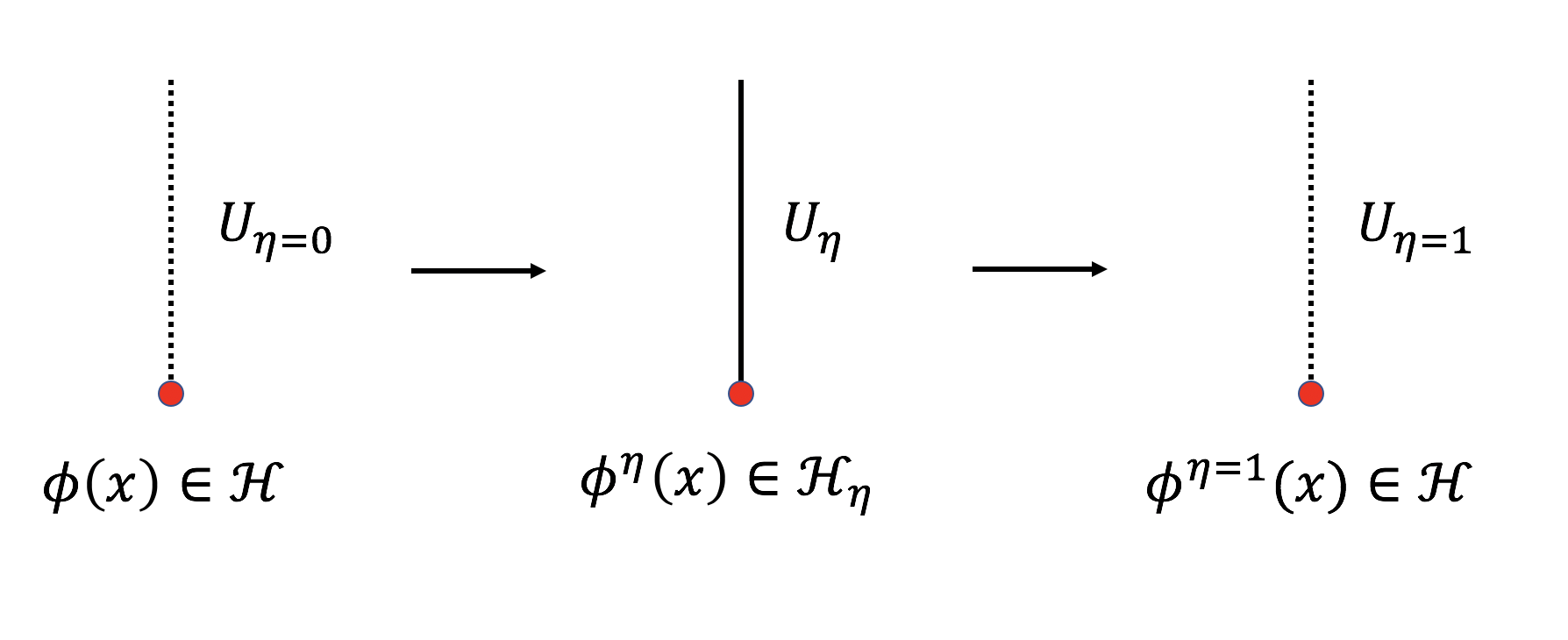}
	\caption{Starting with a local operator $\phi(x)\in {\cal H}$, we can adiabatically turn on a topological line $U_\eta$ from $\eta=0$ (the trivial line, shown in dashed line) to a small but finite value of $\eta$.  This implements the spectral flow map from a local operator $\phi(x)$ to a non-local one $\phi^\eta(x)$ living at the end of the line.  When $\eta$ increases to 1 (corresponding to the $2\pi$ rotation of $U(1)$), the topological line becomes trivial, and we end up with another local operator $\phi^{\eta=1}(x)\in {\cal H}$.  The whole process implements a spectral flow by one unit that maps a local operator $\phi(x)$ to another $\phi^{\eta=1}(x)$.  }\label{fig:adiabatic}
	\end{figure}

	So far we have not used the fact that the underlying symmetry is globally $U(1)$ instead of $\mathbb{R}$. 
	If the global structure is $U(1)$, then the charge $U_\eta$ is trivial when $\eta\in \mathbb{Z}$.  
	In particular, the twisted Hilbert space ${\cal H}_{\eta=1}$ reduces to the Hilbert space $\cal H$ of local operators.  
	In other words, spectral flow with integer units $\eta\in \mathbb{Z}$ maps a local operator to another.  See Figure \ref{fig:adiabatic}. 
	This property would not have held if the global structure of the symmetry is given by $\mathbb{R}$ instead of $U(1)$.

	Let us study the consequences of the $U(1)$ symmetry.  
From  \eqref{sf}, the  operator $\phi^\eta$ obtained from the spectral flow has Lorentz spin 
\ie
L_0^\eta - \bar L_0^\eta  =(  L_0-\bar L_0)- \eta(J_0+\bar J_0)+ {\eta^2\over2} (k-\bar k)\,.
\fe
Now let us assume $\eta\in \bZ$, so the operator $\phi^\eta$ is a local operator.  
	In a bosonic CFT, since the original local operator $\phi$ has integral Lorentz spin and $U(1)$ charge, i.e.\ $L_0-\bar L_0\in \bZ$ and $J_0+\bar J_0\in \mathbb{Z}$, the quantization of the Lorentz spin for $\phi^\eta$  implies that the levels have to obey
	\begin{align}\label{anomaly}
	(\text{bosonic}):~~	k-\bar k \in 2\mathbb{Z}\,,
	\end{align}
	which is the 't Hooft anomaly of the  $U(1)$ global symmetry.   
	If the CFT is fermionic, the Lorentz spin can be half integer and the quantization of the levels is modified to 
	\begin{align}\label{fanomaly}
	(\text{fermionic}):~~	k-\bar k \in \mathbb{Z}\,,
	\end{align}
	
	\subsection{Example: Compact Boson}\label{sec:boson}
	
	Let us illustrate the above general discussions in the example of $c=1$ compact boson  at radius $R$.  
	Let $X(z,\bar z) =X_L(z)+X_R(\bar z)$ be the compact boson field with the identification  $X(z,\bar z) \sim X(z,\bar z)+2\pi R$.  We normalize the OPE to be\footnote{Our convention for the compact boson radius is such that $R=1$ corresponds to the self-dual point with enhanced $\mathfrak{su}(2)$ current algebra.}
	\begin{align}
		X(z,\bar z) X(0,0)\sim -\frac 12 \log |z|^2\,.
	\end{align}
	Hence $\partial X(z) \partial X(0)\sim -{1\over 2z^2}$ and $\bar \partial X(\bar z) \bar\partial X(0)\sim -{1\over 2\bar z^2}$.  The $\mf{u}(1)_k\times$Virasoro  primaries are the exponential operators 
	\ie
	{\cal O}_{n,w}(z,\bar z) = \exp\left[  i\left( {n\over R} +wR\right) X_L(z)  +   i\left( {n\over R} -wR\right) X_R(\bar z)  \right]\,,
	\fe
	with integer momentum $n$ and winding numbers $w$. 
	The conformal weights are
	\ie
	h=\frac 14 \left(  \frac nR +wR\right)^2\,,~~~~\bar h=\frac 14 \left(  \frac nR-wR\right)^2\,.
	\fe
	 We have the OPE
	\begin{align}
		i\partial X(z) {\cal O}_{n,w}(0) \sim { \left( \frac nR +wR\right)\over2 z} {\cal O}_{n,w}(0)\,,~~~
		i\bar\partial X(\bar z) {\cal O}_{n,w}(0) \sim { \left( \frac nR -wR\right)\over2\bar z} {\cal O}_{n,w}(0)\,.
	\end{align}
	For generic $R$, the charges of $i\partial X(z)$ and $i\bar\partial X(\bar z)$ are irrational. This implies that they generate two $\mathbb{R}$ groups instead of $U(1)$.

	At any radius $R$, there are two $U(1)$ global symmetries, the momentum $U(1)_n$ and the winding $U(1)_w$, that are not holomorphic nor antiholomorphic. 
	The currents of the momentum and winding $U(1)$'s are  combinations of $\partial X(z)$ and $\bar \partial X(\bar z)$:
	\begin{align}
		&U(1)_{n} :~~J(z) =  iR  \partial X(z)\,,~~~~~\bar J(\bar z) = i R \bar\partial X(\bar z)\,,\\
		&U(1)_{w} :~~J(z) =  {i\over R}  \partial X(z)\,,~~~~~\bar J(\bar z) =- {i\over R} \bar\partial X(\bar z)\,,
	\end{align}
	under which ${\cal O}_{n,w}$ has integer charges $n$ and $w$, respectively.  The levels $k$ and $\bar k$ of the two $U(1)$'s are
	\begin{align}
		&U(1)_{n} :~~ k =\bar k= { R^2\over2} \,,\\
		&U(1)_{w} :~~ k =\bar k= { 1\over2R^2} \,.
	\end{align}
	In particular, they both obey $k=\bar k$, which means that they are non-anomalous (but there is a mixed anomaly between them).  
	Note that $k$ and $\bar k$ are not separated quantized in general.  
	When $R^2$ is rational, there is an integral linear combination of $U(1)_n$ and $U(1)_w$ that is holomorphic, and another integral combination that is antiholomorphic.  By contrast, at an irrational radius, there is no holomorphic nor antiholomorphic $U(1)$ global symmetry, but only $\mathbb{R}$.

	The  spectral flow \eqref{sf} for $U(1)_{n}$ by one unit $\eta=1$ maps the exponential operator ${\cal O}_{n,w}$ to ${\cal O}_{n,w-1}$, while the spectral flow for  $U(1)_{w}$ maps ${\cal O}_{n,w}$ to ${\cal O}_{n-1,w}$.

	\section{Vanishing Twist Gap}
	
	\subsection{Compactness of Global Symmetry}
	\label{sec:compact}
We will show, under some assumption which we will spell out later, that a current $(J,\bar J)$ acting on a 
CFT with a discrete spectrum generates a compact $U(1) \times U(1)$ symmetry as opposed to $U(1) \times \mathbb R$ 
or $\mathbb R \times \mathbb R$. \footnote{It is also possible that the rank of the symmetry is one. 
This happens, for example, when $\bar k = 0$ and $\bar J$ vanishes identically. In this case, 
a straightforward generalization of the following argument shows that the symmetry group is $U(1)$ rather
than $\mathbb{R}$, and our argument for the vanishing twist gap in Section \ref{sec:nonholo} will follow.  
There is also a logical possibility that $k$ and $\bar k$ are both non-zero, but 
 the left- and right-moving charges are correlated   (i.e. $J_0 = C \bar J_0$ for some nonzero constant $C$) in such a way that the rank of the global symmetry is still one 
(for example, the $c=1$ free boson with only momentum modes and no winding modes). However, this does not happen for a compact CFT. If the rank were one, a straightforward generalization of the following
argument shows that 
the left-moving charges are all rational numbers with a common denominator, which means the left-moving $U(1)$ symmetry is compact.  
Then the spectral flow operators  (\ref{eq:holomorphictower})   of this $U(1)$ (which also follow from modular covariance) would violate  $J_0 = C \bar J_0$, and hence a contradiction. } 
For definiteness, let us first assume that there are no other conserved currents, though our argument can be easily generalized to cases with additional currents.

Let us first look at the spectrum of the left-moving charge $J_0$. Assuming faithfulness, there must be at least one operator with a non-zero charge. Let us normalize $J(z)$ so that the lightest charged state carries $J_0 = 1$. We claim 
that all other charges can be written as either as $J_0=q$ with $q \in \mathbb Q$ or 
$J_0 = q_1 + x q_2$, where $q_1, q_2 \in \mathbb Q$ and $x$ is some irrational number. To show this, 
suppose to the contrary that 
there were a state with $J_0 = y$ for some irrational $y$ that cannot be written as $y = q_1 + x q_2$. 
We would then need at least  three rational numbers 
$q_1, q_2,$ and $q_3$ to parametrize $J_0$ eigenvalues as $J_0 = q_1 + x q_2 + y q_3$. 
The operator product expansion would then preserve $q_1, q_2,$ and $q_3$
separately. Thus, we would conclude that the rank of global symmetry preserving the spectrum and the operator
product expansion is at least three. By the Noether theorem, there must then be at least three conserved currents,
contradicting our assumption that there are no other conserved currents.

We can repeat the argument for the right-mover and conclude that the $\bar J$ spectrum must be 
either $\bar J_0 = p$ with $p \in \mathbb Q$ or  
$\bar J_0 = p_1 + x p_2$ with $p_1, p_2 \in \mathbb Q$. We must have the same irrational number $x$
for both $J$ and $\bar J$,
otherwise there must be at least three conserved currents. Moreover, the spectra of $J_0$ and $\bar J_0$
must be correlated; we must be able to take their two linearly independent combinations
 such that their eigenvalues are both in $\mathbb Q$.
If it is not possible to take such linear combinations, the rank of the global symmetry must be greater than two,
contradicting our assumption.

 We have been able to show that, if a current $(J, \bar J)$ acts faithfully on a CFT and if there are no other 
conserved currents, then there must be two linearly independent combinations of $J_0$ and $\bar J_0$ such that
their eigenvalues are all rational. In order for the these two combinations to generate a compact 
$U(1) \times U(1)$ symmetry, these rational charges must have their denominators
bounded. If the CFT has a discrete spectrum, then for any $\Delta > 0$, there are a finite number of states with
conformal dimensions below $\Delta$ and therefore the denominators are bounded. 
The question is what happens when we take the limit $\Delta \rightarrow \infty$.

In section 6 of \cite{Harlow:2018tng}, Daniel Harlow and one of the authors showed that,
if a CFT spectrum is finitely generated, then any non-compact global symmetry of that CFT is a subgroup of 
a larger compact global symmetry\footnote{Here, we are discussing compactness 
applicable to general CFT not necessarily with a weakly coupled gravity dual.  
For compactness of gauge symmetry in gravitational theory, 
see also the earlier papers \cite{Polchinski:2003bq, Banks:2010zn}.}
 ``Finitely generated'' means that there is a finite subset of the
primary fields whose operator product expansion generates all other primary fields (See
 \cite{Harlow:2018tng} for a precise formulation). In our case, the proof is straightforward. 
Since rational charges of a finite subset of primary fields have bounded denominators, 
so are all other charges generated by the operator product expansion.
This shows that
the global symmetry generated by $(J, \bar J)$ is $U(1) \times U(1)$ or is $U(1)$ when
either $J=0$ or $\bar J=0$.

We hope that the assumption that the CFT spectrum is finitely generated can be relaxed,
at least in two dimensions.

	\subsection{Twist Gap and the $U(1)$ Global Symmetry}
	\label{sec:nonholo}

  Consider a 2d compact, unitary CFT with a $U(1)$ global symmetry generated by the  current $(J,\bar J)$.  
Unitarity and current conservation imply that both components of the current  are conserved, so they separately generate two abelian symmetries.  
	In the previous section, we have argued that the full global symmetry associated to $(J,\bar J)$ is 
either $U(1)\times U(1)$ or $U(1)$. Note that, in the rank two case, each $U(1)$ factor is not necessarily 
holomorphic or antiholomorphic. 
	As reviewed in Section \ref{sec:boson}, this is the case for the $c=1$ free boson at a generic radius, where  the left and right moving charges with respect to $J$ and $\bar J$ are irrational, while the momentum and winding charges are quantized.

As a warmup, let us show 
that the twist gap is zero for the special case when the symmetry group is $U(1)$ and
 is generated by  a holomorphic current  or an antiholomorphic current.
	Let us assume the $U(1)$ current is holomorphic, while the antiholomorphic case works identically. 
To avoid confusion, we denote the abelian current algebra generated by the modes $J_{n}$ of a holomorphic   current $J$ with level $k$ by $\mathfrak{u}(1)_k$. 
	If the holomorphic global symmetry is globally $U(1)$, then $\bar k =0$ and the level has to be quantized as $k\in 2\mathbb Z$ from \eqref{anomaly}.  
	In this case, the spectral flow by $\pm1$ unit \eqref{sf} maps the identity operator to to a pair of local operators with quantum number
	\ie
	&L_0= {k\over2} \,,~~~J_0 = \mp k \,,\\
	&\bar L_0 = 0\,,~~~\bar J_0=0\,.
	\label{eq:holomorphictower}
	\fe
Therefore, the current algebra is enhanced by  these higher spin currents.  
These (higher-spin) spectral flow currents are necessarily $\mathfrak{u}(1)_k$ primaries.\footnote{We prove this statement below. Suppose the spectral flow current associated to $\eta=1$ is an $\mathfrak{u}(1)_k$ descendant  of a primary $\cal O$ with $J_0=-k$ and $L_0<k/2$.   Then we can apply another  spectral flow with $\eta=-1$ on $\cal O$ to obtain a local operator with $J_0=0$ and $L_0<0$, which would violate  unitarity.} Consequently, any compact CFT with holomorphic $U(1)$ symmetry has a tower (and the complex conjugate) of $\mf{u}(1)_k$ primaries of zero twist from the spectral flow currents. 

It remains to show that the twist gap is zero when the holomorphic current generates an $\mathbb R$ symmetry instead of $U(1)$.
	
	From Section \ref{sec:compact}, we conclude that the global form of the symmetry is $U(1)\times U(1)$.
	By assumption, one of the $U(1)$ symmetries is generated by the (non-holomorphic) current $(J,\bar J)$.  
Let us assume the other $U(1)$ is generated by the current $(\A J,\B \bar J)$ for some   real numbers $\A,\B$ (with $\A\neq\B$).
 Importantly,  we assume  $\alpha, \beta$ are irrational numbers. Otherwise an integer linear combination of the two compact $U(1)$'s would generate a holomorphic  $U(1)$ symmetry and our previous argument would be sufficient to show vanishing twist gap. 
The levels of the second $U(1)$ are ($\alpha^2 k, \beta^2 \bar k)$.  
	For each   $U(1)$ factor, we can perform independent spectral flows as explained in Section~\ref{sec:spectralflow}.\footnote{For the $c=1$ free boson at radius $R$, if we take the momentum $U(1)_n$ to be the first non-holomorphic $U(1)$, then $k=\bar{k} = \frac{R^2}2, \alpha = \frac1{R^2}, \beta = -\frac{1}{R^2}$ (see Section \ref{sec:boson}).} 
	
	With respect to the first $U(1)$ symmetry, spectral flow   by $\eta \in \mathbb Z$ units takes a state in the Hilbert space $\cal H$ to another with the following changes in  quantum numbers given by (\ref{sf}).
	Meanwhile spectral flow with respect to the second $U(1)$ by $\eta' \in \mathbb Z$ units induces
	\begin{align} 
		L_0 &\rightarrow L_0 - \eta' \alpha J_0 + \frac{\eta'^2 \alpha^2 k}2 \nn\\
		\alpha J_0 &\rightarrow \alpha J_0 -  \eta' \alpha^2 k \nn\\
		\bar L_0 &\rightarrow \bar L_0  + \eta' \beta \bar J_0  + \frac{\eta'^2 \beta^2 \bar k}2 \nn\\
		\beta \bar J_0 &\rightarrow \beta \bar J_0 + \eta' \beta^2 \bar k
		\label{sp2}
	\end{align}
	which follows from  \eqref{sf} with the replacement $J_0 \to \alpha J_0$, $k \to \alpha^2 k$, $\bar{J_0} \to \beta \bar{J_0}$, $\bar k \to \beta^2 \bar k$. 
	
	Let us now take the vacuum state and spectral flow first by $\eta \in \mathbb Z$ units with respect to the first $U(1)$, and $\eta' \in \mathbb Z$ units  with respect to  the second $U(1)$ (or vice versa). The resulting state will have quantum numbers 
	\begin{align}
		L_0^{\eta, \eta'} &= \frac{k}{2}\(\eta+\eta'\alpha\)^2\nn\\
		\bar L_0^{\eta,\eta'} &= \frac{\bar k}2\(\eta+\eta'\beta\)^2 \nn\\
		J_0^{\eta, \eta'} &= -k(\eta + \eta'\alpha) \nn\\
		\bar{J_0}^{\eta, \eta'} &= \bar k(\eta+\eta'\beta).
	\end{align} 
	By choosing the integers $\eta, \eta'$, we can have $\eta+\eta'\alpha$ arbitrarily close to zero.\footnote{Here we have used Dirichlet's approximation theorem, which states that for any irrational number $\alpha$, the inequality $|\alpha+ \eta/\eta'|<{1\over (\eta')^{2}}$ is satisfied for infinitely many integers $\eta,\eta'$. We thank Petr Kravchuk and Juan Maldacena for discussions on this point.}  Hence we have shown that the twist gap of this theory must indeed vanish.
	
	Note that \eqref{sp2} implies the following conditions on the current levels from spin quantization (invariance of the theory under modular $T$ transformations):
	\begin{align}
		\alpha k - \beta\bar{k} &\in \mathbb{Z} \nn\\
		\alpha^2 k - \beta^2\bar{k} &\in 2\mathbb{Z}.
\label{eq:quant}
	\end{align}
These theories all come with an exactly marginal operator, $J \bar J$. The quantization conditions (\ref{anomaly}) and (\ref{eq:quant}) imply that under deformation by this operator, the parameters $\alpha, \beta, k, \bar k$ must change in a controlled way. In particular, if 
\begin{align}
\alpha \rightarrow \alpha + \delta \alpha \,,~~~\beta \rightarrow \beta + \delta \beta \,,~~~
k \rightarrow k + \delta k \,,~~~\bar k \rightarrow \bar k + \delta \bar k
\label{eq:def}
\end{align}
under a small deformation, then 
	\be \delta k = \delta \bar k, ~~~~~~~~\delta \alpha = \frac{\beta-\alpha}{2k}\delta k, ~~~~~~~ \delta \beta = \frac{\alpha-\beta}{2\bar k}\delta k, \label{eq:inf} \ee where the first condition comes from \eqref{anomaly}. It can be checked that the free boson under a small deformation in $R$ obeys (\ref{eq:inf}).
	
We pause here to note that we have two rather qualitatively different arguments to show the twist gap vanishes for all 2d CFTs with an abelian current. The first is the holomorphic current $J$ is associated with a compact global symmetry $U(1)$. In this case, the spectral flow operation on the vacuum state maps to a new charged state with twist $0$ that is not a vacuum descendent. The second is if the global $U(1)\times U(1)$ is generated by currents $(J, \bar J)$ and $(\alpha J, \beta \bar J)$ for irrational $\alpha, \beta$. In this case, we cannot show there must be a nontrivial primary of zero twist in the CFT, but we can show there is an infinite tower of states that accumulate to zero twist.

The latter situation that we described in the above paragraph is the generic situation for 2d CFTs with an abelian current. If the global $U(1)\times U(1)$ symmetry is generated by currents $(J, \bar J)$ and $(\alpha J, \beta \bar J)$ for at least one of $\alpha, \beta$ rational, then we can simply deform by the $J\bar{J}$ operator which would change $\alpha, \beta$ as in (\ref{eq:inf}) and generically make them both irrational. (For example, $\alpha,\beta$ are generically irrational in the $c=1$ compact boson.)
This means that a generic 2d CFT with an abelian current has an accumulation point at zero twist. 
Since these operators have arbitrarily small but nonzero twist, they cannot be the vacuum descendants of any chiral algebra.  
		
We end this section by remarking that compactness of the $U(1)$ symmetry is essential in our argument for a vanishing twist gap. The crossing equation (\ref{holoS}) alone is not enough to prove a vanishing twist gap. As a counterexample, we can take the $\mf{u}(1)_k\times$Virasoro vacuum character and perform a regularized sum over modular images in the same fashion as \cite{Maloney:2007ud, Keller:2014xba}. This will give a finite, modular-invariant spectrum with a unique $PSL(2,\mathbb C)$-invariant vacuum, and a continuous energy spectrum at each spin, with continuous charges. From the modular kernels of the $\mf{u}(1)_k\times$Virasoro characters (see Appendix C of \cite{Benjamin:2016fhe}), the spectrum will only have support when the twist is at least $\frac{c-2}{12}$.\footnote{Here we assumed $c>2$. In general the vacuum $\mf{u}(1)_k\times$Virasoro character is the product the of $\mf{u}(1)_k$ vacuum character $\chi^{\mf{u}(1)_k}_0={q^{1\over 24}\over \eta(q)}$, and the Virasoro vacuum character $\chi^{\rm Vir}_0$ at central charge $c-1$  \cite{Feigin, RochaCaridi}. For $c>2$, under S-transformation, the modular kernel has support for $h\geq {c-2\over 24}$ as shown in \cite{Benjamin:2016fhe}. However for $c\leq 2$, the modular S-transformation also has support at the vacuum $h=0$ (since the the S-transformation of $\chi^{\rm Vir}_0$ does). Consequently, such $c<2$ CFTs with $\mf{u}(1)$ symmetry necessarily have twist zero states (higher-spin currents).
}
This means the lightest charged states have a  finite twist gap. Since the charge is continuous, this theory (if it exists) does not have  compact $U(1)$ symmetries and is not finitely generated, and therefore does not contradict our result.
	
	\subsection{Applications to ${\cal N}=2$ Theories}
	\label{sec:n2}
	
	We now apply the results of Sections \ref{sec:compact} and \ref{sec:nonholo} to the case of $\mathcal{N}=(2,2)$ superconformal field theories with $c>3$. Recall that the unextended $\cN=2$ super-Virasoro algebra contains  as generators, the supercurrents $G^\pm(z)$, R-current $J_R(z)$ in addition to the stress tensor $T(z)$. We would like to show that any compact $c>3$ $\cN=(2,2)$ SCFT  has a vanishing twist gap with respect to the unextended $\cN=2$ super-Virasoro algebra.  
 When $c<3$, there are $\cN=2$ minimal models whose twist gaps are nonzero.

There are two cases to consider:  the holomorphic R-symmetry is globally $U(1)$ or $\mathbb{R}$. We start with the case where the holomorphic R-symmetry is $U(1)$.  
	Let the holomorphic level of the current $J_R(z)$ be $k$, normalized such that   the $U(1)$ R-symmetry acts faithfully with all the states having integer charges.  
	In particular, the level $k$ is the 't Hooft anomaly \eqref{fanomaly} of the $U(1)$ R-symmetry and has to be an integer, $k\in \bZ$. 
	This is different from the usual convention where the supercurrents  have unit charges
		\ie
	q_{\rm usual}(G^\pm)=\pm 1,
	\fe
	while the charged operators might have fractional charges.  
	Instead, we normalize the charges so that the supercurrents have
	\ie\label{ourq}
	q(G^\pm)=\pm \sqrt{\frac{3k}c} \in \mZ.
	\fe Accordingly, the BPS condition in the NS sector is now
	\be
	h = \sqrt{\frac{c}{12k}}{q}
	\ee
	rather than the more familiar $h={1\over 2}{q_{\rm usual}} $.
	
	When the holomorphic R-symmetry is globally $U(1)$, we can apply   an integer unit of spectral flow on the identity operator to obtain
	\be
	h = \frac{k  }2, ~~~\bar h =0, ~~~ q =  k,~~~ \bar q =0,
	\label{eq:SFstates}
	\ee
 similarly if we start with any other chiral operator in the $\cN=2$ vacuum multiplet. 
	To prove that the twist gap with respect to the $\cN=2$ Virasoro algebra is zero, it remains to show that this type of spectral flowed operator is not an $\cN=2$ descendant of the identity when $c>3$.    
	
	We will prove this statement by contradiction.  
	Suppose that every spectral flow image of general chiral generators of the $\cN=2$ algebra  is a $\cN=2$ vacuum descendant.  
	The identity character $\chi_0^{\text{NS}}$ must be invariant under spectral flow, which requires
	\be
	\chi_0^{\text{NS}}(\tau, z) = \chi_0^{\text{NS}}(\tau, z + \tau) y^k q^{\frac k2}.
	\label{sfid}
	\ee
	For theories with $c>3$, the NS sector vacuum character is  given by \cite{Boucher:1986bh, Dobrev:1986hq, Kiritsis:1986rv, Eguchi:1988af}
	\be
	\chi_0^{\text{NS}}(\tau, z) = q^{-\frac{c-3}{24}} \frac{1-q}{(1+y^{\sqrt{\frac{3k}{c}}}q^{1/2})(1+y^{-\sqrt{\frac{3k}{c}}}q^{1/2})}\frac{\theta_3(\tau,\sqrt{\frac{3k}{c}}z)}{\eta(\tau)^3}.
	\label{eq:vacchar}
	\ee
	Recall $\sqrt{3k/c}$ is the (normalized) $U(1)_R$ charge of the supercurrent $G^+$ which is required to be a positive integer. Then \eqref{sfid} simply requires
	\ie
	q^{-{k\over 2}}y^{-k}={(1+y^{\sqrt{3k\over c}}q^{{1\over 2}+\sqrt{3k\over c}})(1+y^{-\sqrt{3k\over c}}q^{{1\over 2}-\sqrt{3k\over c}})\over (1+y^{\sqrt{3k\over c}}q^{1/2})(1+y^{-\sqrt{3k\over c}}q^{{1\over 2}})}{\theta_3(\tau,\sqrt{3k\over c}(z+\tau))\over \theta_3(\tau, \sqrt{3k\over c} z)}. 
	\fe 
	Let us expand both sides around small $q$. Recall that $\theta_3(\tau,z)=\sum_{n\in \bZ} q^{n^2/2}y^n$, thus $\theta_3(\tau,\A(z+\tau))$ is dominated by the summand with $n=-\A$ and gives (with $\A=\sqrt{3k\over c}$ here),
	\ie
	q^{-{k\over 2}}y^{-k}=q^{- {3k\over 2c} }y^{-  {3k\over c}  }
	\fe
	which  is not possible for $c>3$.  

	This implies that if the holomorphic R-symmetry is $U(1)$, all $\mathcal{N}=2$ SCFTs with $c>3$ have vanishing twist gaps with respect to the un-extended $\cN=2$ algebra.\footnote{For the $\cN=2$ minimal models whose central charge are $c=\frac{3m}{m+2}$, the vacuum character is different, and is indeed invariant under spectral flow for $k=4m(m+2)$ or $m(m+2)$ for odd and even $m$ respectively.  Correspondingly, the spectral flow currents \eqref{eq:SFstates} are $\cN=2$ descendants of the identity, and the twist gap is nonzero.}

When $c=3\hat c$ is a multiple of 3 (i.e. $\hat c\in \bZ$) and when $k=c/3$, the (unextended) ${\cal N}=2$ Virasoro algebra is extended by the spectral flow current \eqref{eq:SFstates} which has $h=\hat c/2$. The spectral flow current and its superpartner, together with the $\mathcal{N}=2$ Virasoro algebra, generate the extended ${\cal N}=2$ Virasoro algebra \cite{Odake:1988bh}. This is the algebra of, for instance, sigma models with target space Calabi-Yau manifold.

As a corollary, for any $\mathcal{N}=(2,2)$ SCFT whose holomorphic R-symmetry is $U(1)$ (as opposed to $\mathbb R$), the quantization of the R-charge of the supercurrent \eqref{ourq} and that of the level $k\in \bZ$ \eqref{fanomaly} imply that the central charge must be rational. 
In other words, in $\mathcal{N}=(2,2)$ SCFTs with irrational central charges\footnote{In \cite{Benini:2013cda}, a large class of 2d SCFTs are constructed from compactifications of 4d and 6d SCFTs and their central charges are determined by  $c$-extremization. However all examples considered there with  $\cN=(2,2)$ supersymmetry  turn out to have rational central charges.}, the  holomorphic R-symmetry must globally be  $\mathbb R$ (as opposed to $U(1)$).

We are now left with  the case when the holomorphic R-symmetry is $\mathbb{R}$.  
 The argument in Sections \ref{sec:compact} still implies that the global form of the R-symmetry group is  $U(1)\times U(1)$, but each factor of $U(1)$ is not generated by a holomorphic or an antiholomorphic current. 
As in Section \ref{sec:nonholo}, we apply integer units of spectral flows for both (non-holomorphic) $U(1)$'s on the identity to generate a sequence of states with arbitrarily low twists.  Since these states have arbitrarily small but nonzero twists, they cannot be $\cN=2$ descendants of the identity. Therefore the twist gap with respect to the (unextended) $\cN=2$ Virasoro algebra vanishes even without assuming the holomorphic R-symmetry is globally $U(1)$.

\section*{Acknowledgements}
We thank Igor Klebanov, Petr Kravchuk, Ying-Hsuan Lin, Juan Maldacena, and Xi Yin for interesting discussions.
The work of N.B. is supported in part by the Simons Foundation Grant No. 488653.
The work of H.O. is supported in part by
U.S.\ Department of Energy grant DE-SC0011632,
by the World Premier International Research Center Initiative,
MEXT, Japan,
by JSPS Grant-in-Aid for Scientific Research C-26400240,
and by JSPS Grant-in-Aid for Scientific Research on Innovative Areas
15H05895.
The work of S.H.S. is supported by the Simons Collaboration on Ultra-Quantum Matter, which is a grant from the Simons Foundation (651440, NS).  
The work of Y.W. is  supported in part by the US NSF under Grant No. PHY-1620059 and by the Simons Foundation Grant No. 488653. 
We thank the Aspen Center for Theoretical Physics, which is supported by
the National Science Foundation grant PHY-1607611,  where this work was initiated.

	\bibliographystyle{JHEP}
	\bibliography{U1}

\providecommand{\href}[2]{#2}\begingroup\raggedright\begin{thebibliography}{10}

\bibitem{Cardy:1986ie}
J.~L. Cardy, {\it {Operator Content of Two-Dimensional Conformally Invariant
  Theories}},  {\em Nucl. Phys.} {\bf B270} (1986) 186--204.

\bibitem{Collier:2016cls}
S.~Collier, Y.-H. Lin, and X.~Yin, {\it {Modular Bootstrap Revisited}},  {\em
  JHEP} {\bf 09} (2018) 061, [\href{http://arxiv.org/abs/1608.06241}{{\tt
  arXiv:1608.06241}}].

\bibitem{Benjamin:2019stq}
N.~Benjamin, H.~Ooguri, S.-H. Shao, and Y.~Wang, {\it {Light-cone modular
  bootstrap and pure gravity}},  {\em Phys. Rev.} {\bf D100} (2019), no.~6
  066029, [\href{http://arxiv.org/abs/1906.04184}{{\tt arXiv:1906.04184}}].

\bibitem{Alday:2019vdr}
L.~F. Alday and J.-B. Bae, {\it {Rademacher Expansions and the Spectrum of 2d
  CFT}},  \href{http://arxiv.org/abs/2001.00022}{{\tt arXiv:2001.00022}}.

\bibitem{Bhardwaj:2017xup}
L.~Bhardwaj and Y.~Tachikawa, {\it {On finite symmetries and their gauging in
  two dimensions}},  {\em JHEP} {\bf 03} (2018) 189,
  [\href{http://arxiv.org/abs/1704.02330}{{\tt arXiv:1704.02330}}].

\bibitem{Chang:2018iay}
C.-M. Chang, Y.-H. Lin, S.-H. Shao, Y.~Wang, and X.~Yin, {\it {Topological
  Defect Lines and Renormalization Group Flows in Two Dimensions}},  {\em JHEP}
  {\bf 01} (2019) 026, [\href{http://arxiv.org/abs/1802.04445}{{\tt
  arXiv:1802.04445}}].

\bibitem{Lin:2019kpn}
Y.-H. Lin and S.-H. Shao, {\it {Anomalies and Bounds on Charged Operators}},
  {\em Phys. Rev.} {\bf D100} (2019), no.~2 025013,
  [\href{http://arxiv.org/abs/1904.04833}{{\tt arXiv:1904.04833}}].

\bibitem{Schwimmer:1986mf}
A.~Schwimmer and N.~Seiberg, {\it {Comments on the N=2, N=3, N=4 Superconformal
  Algebras in Two-Dimensions}},  {\em Phys. Lett.} {\bf B184} (1987) 191--196.

\bibitem{Benjamin:2016fhe}
N.~Benjamin, E.~Dyer, A.~L. Fitzpatrick, and S.~Kachru, {\it {Universal Bounds
  on Charged States in 2d CFT and 3d Gravity}},  {\em JHEP} {\bf 08} (2016)
  041, [\href{http://arxiv.org/abs/1603.09745}{{\tt arXiv:1603.09745}}].

\bibitem{Harlow:2018tng}
D.~Harlow and H.~Ooguri, {\it {Symmetries in quantum field theory and quantum
  gravity}},  \href{http://arxiv.org/abs/1810.05338}{{\tt arXiv:1810.05338}}.

\bibitem{Polchinski:2003bq}
J.~Polchinski, {\it {Monopoles, duality, and string theory}},  {\em Int. J.
  Mod. Phys.} {\bf A19S1} (2004) 145--156,
  [\href{http://arxiv.org/abs/hep-th/0304042}{{\tt hep-th/0304042}}].
  [,145(2003)].

\bibitem{Banks:2010zn}
T.~Banks and N.~Seiberg, {\it {Symmetries and Strings in Field Theory and
  Gravity}},  {\em Phys. Rev.} {\bf D83} (2011) 084019,
  [\href{http://arxiv.org/abs/1011.5120}{{\tt arXiv:1011.5120}}].

\bibitem{Maloney:2007ud}
A.~Maloney and E.~Witten, {\it {Quantum Gravity Partition Functions in Three
  Dimensions}},  {\em JHEP} {\bf 02} (2010) 029,
  [\href{http://arxiv.org/abs/0712.0155}{{\tt arXiv:0712.0155}}].

\bibitem{Keller:2014xba}
C.~A. Keller and A.~Maloney, {\it {Poincare Series, 3D Gravity and CFT
  Spectroscopy}},  {\em JHEP} {\bf 02} (2015) 080,
  [\href{http://arxiv.org/abs/1407.6008}{{\tt arXiv:1407.6008}}].

\bibitem{Feigin}
B.~L. Feigin and D.~B. Fuchs {\em Lecture Notes in Math.} {\bf 1060} (1984)
  230--245.

\bibitem{RochaCaridi}
A.~Rocha-Caridi {\em Vertex Operators in Mathematics and Physics, eds. J.
  Lepowsky, S. Mandelstam, I. M. Singer,} (1985) 451--473.

\bibitem{Boucher:1986bh}
W.~Boucher, D.~Friedan, and A.~Kent, {\it {Determinant Formulae and Unitarity
  for the N=2 Superconformal Algebras in Two-Dimensions or Exact Results on
  String Compactification}},  {\em Phys. Lett.} {\bf B172} (1986) 316.

\bibitem{Dobrev:1986hq}
V.~K. Dobrev, {\it {Characters of the Unitarizable Highest Weight Modules Over
  the $N=2$ Superconformal Algebras}},  {\em Phys. Lett.} {\bf B186} (1987) 43.

\bibitem{Kiritsis:1986rv}
E.~Kiritsis, {\it {Character Formulae and the Structure of the Representations
  of the $N=1$, $N=2$ Superconformal Algebras}},  {\em Int. J. Mod. Phys.} {\bf
  A3} (1988) 1871.

\bibitem{Eguchi:1988af}
T.~Eguchi and A.~Taormina, {\it {On the Unitary Representations of $N=2$ and
  $N=4$ Superconformal Algebras}},  {\em Phys. Lett.} {\bf B210} (1988)
  125--132.

\bibitem{Odake:1988bh}
S.~Odake, {\it {Extension of $N=2$ Superconformal Algebra and Calabi-yau
  Compactification}},  {\em Mod. Phys. Lett.} {\bf A4} (1989) 557.

\bibitem{Benini:2013cda}
F.~Benini and N.~Bobev, {\it {Two-dimensional SCFTs from wrapped branes and
  c-extremization}},  {\em JHEP} {\bf 06} (2013) 005,
  [\href{http://arxiv.org/abs/1302.4451}{{\tt arXiv:1302.4451}}].

\end{thebibliography}\endgroup
	
\end{document}